# Observation of large spin accumulation voltages in non-degenerate Si spin devices due to spin drift effect: Experiments and theory

**Takayuki Tahara [1,†], Yuichiro Ando[1,†,*], Makoto Kameno[2], Hayato Koike[3], Kazuhito Tanaka[2], Shinji Miwa[2], Yoshishige Suzuki[2], Tomoyuki Sasaki[3], Tohru Oikawa[3], Masashi Shiraishi[1,2,#]**

1. Department of Electronic Science and Engineering, Kyoto University, Japan
2. Graduate School of Engineering Science, Osaka University, Japan.
3. Technology HQ, TDK Corporation, Japan.

\* Corresponding author: Yuichiro Ando(ando@kuee.kyoto-u.ac.jp)

\# Corresponding author: Masashi Shiraishi (mshiraishi@kuee.kyoto-u.ac.jp)

[†]These authors contributed equally

Abstract

**A large spin-accumulation voltage of more than 1.5 mV at 1 mA, i.e., a magnetoresistance of 1.5 Ω, was measured by means of the local three-terminal magnetoresistance in nondegenerate Si-based lateral spin valves (LSVs) at room temperature. This is the largest spin-accumulation voltage measured in semiconductor-based LSVs. The modified spin drift-diffusion model, which successfully accounts for the spin drift effect, explains the large spin-accumulation voltage and significant bias-current-polarity dependence. The model also shows that the spin drift effect enhances the spin-dependent magnetoresistance in the electric two terminal scheme. This finding provides a useful guiding principle for spin metal-oxide semiconductor field-effect transistor (MOSFET) operations.**

**I. Introduction**

Silicon (Si) spintronics is becoming a pivotal field in semiconductor spintronics [1-4]. From the viewpoint of industrial applications, Si spintronics devices have good compatibility with existing Si-LSI technologies and a great advantage over other semiconductor spintronic devices [5-13]. Several Si-based spintronics devices have been proposed [14-18], and in particular, spin metal-oxide-semiconductor field-effect transistors (MOSFET) have attracted much attention because they enable a reconfigurable function in logic circuits with low energy consumption and high integration and overcome the physical limitation of Moore's law. Si is a light element that possesses spatial inversion symmetry and low concentrations of isotopes with non-zero nuclear spins, which enables good spin coherence.

Much effort has been paid to transporting spin polarized current and pure spin current in Si [19-23]. In an early stage of Si spintronics, most of the spin transport studies were limited to using degenerate n- and p-type Si [24-27], because the so-called conductance mismatch problem can present a significant obstacle for injecting spins in non-degenerate Si [28, 29]. However, a metallic conduction state in degenerate Si impedes a gate-voltage-induced modulation of spin and charge transport. Therefore, a realization of spin transport in non-degenerate Si was eagerly awaited. Ballistic spin injection using the Si-based hot-electron spin transistor was used to bypass the above-mentioned problem, and has allowed spin injection into intrinsic Si [19, 30, 31] and non-degenerate n-Si with a doping concentration up to $10^{15}$-$10^{16}$ cm$^{-3}$ [32]. However, the operating temperature of the hot-electron transistors did not reach room temperature (RT), and the intensity of the spin signal was not sufficiently great, although the gate-tunable spin signals have been successfully observed in the similar device using intrinsic Si at low temperature [33]. Thus, an experimental demonstration of a spin MOSFET at RT [34, 35] was a notable milestone in Si spintronics. In the operation of the Si-based spin MOSFET, spin transport in nondegenerate n-Si (with a doping concentration of $2\times10^{18}$ cm$^{-3}$) was electrically realized at RT, and the spin signals were modulated by a gate-voltage application. However, the magnitude of the spin signal was still small, thus representing the next obstacle to overcome in the further development of Si-based spin MOSFETs.

In this paper, we report an observation of large spin accumulation voltages up to 1.5 mV (magnetoresistance of 1.5 Ω) at RT in nondegenerate n-Si, as measured by local three-terminal magnetoresistance (L-3T MR). This value is more than an order of magnitude larger than those in the semiconductor-based lateral spin valves reported to date. The spin accumulation signals show a significant bias-current-polarity dependence, i.e., large spin accumulation voltages only at the positive charge current, and nonlinear current dependence, which is unexplained in the framework of the conventional spin diffusion model established in the metal-based lateral spin valves (LSVs) [36-42]. An expansion of the theory of spin transport that accounts for the spin drift effect in a nondegenerate semiconductor [43, 44] quantitatively reproduces the experimental results. The modified theory successfully explains the bias-current-polarity dependence of the spin accumulation voltage not only in a nondegenerate Si-based LSV but also in a degenerate Si-based and a nonmagnetic Cu-based one. Furthermore, the theory shows a significant enhancement of the magnetoresistance in the electrical two-terminal scheme, which is the typical scheme for operating the spin MOSFETs.

## II. Device fabrication procedure

The Si-based LSV was fabricated on a silicon-on-insulator substrate with 100-nm-thick Si(100) / 200-nm-thick $SiO_2$ / bulk Si(100) (see Fig. 1(a)). The upper Si layer was phosphorous (P) doped by ion implantation. Through a 4-terminal method, the conductivity of the nondegenerate Si channel was determined to be $2.3 \times 10^3$ $\Omega^{-1} m^{-1}$, indicating that the dopant concentration was approximately $2 \times 10^{18}$ $cm^{-3}$. To reduce the contact resistance, a 20-nm-thick highly doped Si layer with a phosphorous concentration of approximately $5 \times 10^{19}$ $cm^{-3}$ was fabricated using ion implantation. The reduction in the interface resistance allows the efficient detection of spin signals. After etching the natural oxide layer on the Si channel using a HF solution, Pd (3 nm) / Fe (13 nm) / MgO (0.8 nm) was grown by molecular beam epitaxy. Then, the Pd (3 nm) / Fe (3 nm) layers were etched, and Ta (3 nm) was grown on the remaining Fe. To form ferromagnetic metal electrodes (FEs), the Si channel was etched to a depth of 25 nm by $Ar^+$ ion milling. The contacts had dimensions of $0.5 \times 21$ $\mu m^2$ and $2 \times 21$ $\mu m^2$. The Si channel surface and sidewalls at the FE1 and FE2 were buried by $SiO_2$. The nonmagnetic electrodes, with dimensions of $21 \times 21$ $\mu m^2$, were made from Al. The gap between the FEs, $d$, was varied from 1.85 to 2.75 $\mu m$. Although the degenerate Si-based LSVs were fabricated in almost the same manner as nondegenerate ones, the thickness of the MgO layer was 1.6 nm. The Si channel with a phosphorous concentration of approximately $5 \times 10^{19}$ $cm^{-3}$ was fabricated by ion implantation. Cu-based LSVs were fabricated on non-doped Si(100) substrates. Two $Ni_{80}Fe_{20}$ (Py) wires with a thickness of 25 nm and a width of 500~1000 nm were fabricated by means of the lift-off process with electron beam lithography and electron beam evaporation. After cleaning the surface of the Py wires by means of a low acceleration $Ar^+$ ion gun, a Cu channel with a thickness of 100 nm and a width of 1500 nm was deposited by electron beam evaporation. All magnetoresistance measurements were carried out at RT

**III. Results**

A schematic of the electrochemical potential of up and down spins in a FE / nonmagnetic channel (NC) / FE structure under an antiparallel configuration is shown in Fig. 1(b). In the local two-terminal magnetoresistance, a change in $|D_V+E_V|/I$ ($I$ : charge current) is detected when a reversal in magnetization-configuration occurs. Although the local two-terminal magnetoresistance is a typical configuration for practical semiconductor-based spin devices, it generally increases the noise level because of the large resistance. Recently, L-3T MR, which also demonstrates the spin transport, has been reported [45]. In the L-3T MR, $I$ is applied between FE2-FE1, whereas the voltage is measured between FE1-NE1, as shown in Fig. 1(a). Namely, the voltage drop only at the FE1 is measured. The advantage of the L-3T MR over the local two-terminal MR that it reduces the noise level, as discussed in section IV(c). Note that the L-3T MR is indeed different from the three-terminal Hanle effect measurement, in which only one ferromagnetic contact is utilized to demonstrate a spin accumulation in an NC, and several spurious signals are often detected simultaneously [46-50].

Room temperature measurements of the L-3T MR at $I=1$ mA are shown in Fig. 2(a). The FE1 contact was placed under the spin extraction condition at positive charge current. A clear spin accumulation signal with steep resistance changes at $H=\pm100$ and $\pm300$ Oe was obtained under the spin extraction condition, as shown in the upper panel of Fig. 2(a). The amplitude of the resistance (voltage) change, $\Delta E_V/I$ ($\Delta E_V$), is approximately 1.5 Ω (1.5 mV), which is more than an order of magnitude larger than those reported in the semiconductor-based LSVs to date. Such a large signal was reproduced in several devices (see Fig. 3). In contrast, although the same contact (FE1) was utilized for spin detection, the amplitude of the hysteresis signal measured under the spin injection condition ($I = -1$ mA) was considerably small, and the hysteresis shape was triangular, not rectangular, as shown in the lower panel of Fig. 2(a). Whereas a small resistance change (0.1 Ω) is recognized at $H=\pm300$ Oe, those at $H=\pm100$ Oe are considerably smaller (< 0.01 Ω). Because the L-3T MR is a local scheme, several spurious signals, e.g., the anisotropic magnetoresistance (AMR), anisotropic tunneling magnetoresistance (TAMR), and magnetoresistance due to the Lorentz force, can be detected [51-56]. Therefore, the triangular hysteresis signal is attributable to the AMR in the Fe layer or TAMR at the Fe/MgO/Si interfaces, and the actual amplitude of the spin accumulation resistance (voltage),

$\Delta D_V/I$ ($\Delta D_V$), is estimated to be less than 0.01 Ω (10 μV). A similarly significant bias-current-polarity dependence, i.e., $\Delta E_V \gg \Delta D_V$, was also reproduced when the FE2 was used for spin detection. Such behavior cannot be explained by the conventional spin drift-diffusion model [36-42]. To confirm that the large rectangular signal can be ascribed to spin accumulation in the Si channel, the Hanle effect measurement in the L-3T scheme was also carried out. The results of the Hanle measurement under parallel and antiparallel configurations are shown in the inset of Fig. 2(b). Whereas large broad peak features due to the spurious magnetoresistance effect overlapped with the spin signals [57-62], the peak and dip features due to the Hanle effect were clearly recognized in the small magnetic field, $|H| \leqq 200$ Oe, under antiparallel and parallel configuration, respectively. It should be noted that the difference in voltage between the parallel and antiparallel configurations at $H = 0$ Oe is comparable with $\Delta E_V$ in Fig. 2(a), indicating that the large rectangular signal is due to the spin accumulation in the Si channel. To estimate the spin lifetime, we used the difference in the Hanle signal between the antiparallel and parallel configurations, as shown in the main panel of Fig. 2(b). In the analysis, we used the following fitting function: [34, 50]

$$\frac{V_{AP}(B)-V_P(B)}{I} = \frac{P^2\sqrt{D\tau_{dr}}}{\sigma A}(1+\omega^2\tau_{dr}^2)^{-\frac{1}{4}}exp\left\{\frac{d}{2\lambda_N^2}v\tau - \frac{d}{\lambda_N}\sqrt{\frac{\sqrt{1+\omega^2\tau_{dr}^2}+1}{2}}\right\}\left\{cos\left(\frac{arctan(\omega\tau_{dr})}{2}\right) + \frac{d}{\lambda_N}\sqrt{\frac{\sqrt{1+\omega^2\tau_{dr}^2}-1}{2}}\right\}, \ldots(1)$$

where $P$ is the spin polarization of the injected current, $A$ is the cross-sectional area of the channel, $D$ is the spin diffusion constant, $\tau$ is the spin lifetime, $\omega = g\mu_B B/\hbar$ is the Larmor frequency, $g$ is the $g$-factor for the electrons ($g = 2$ in this study), $\mu_B$ is the Bohr magneton, $\hbar$ is the Dirac constant, and $v$ is the spin drift velocity. The spin diffusion length is given by $\lambda_N=(D\tau)^{0.5}$. Under spin drift, $\tau_{dr}^{-1} = \frac{v^2}{4D} + \frac{1}{\tau}$. Eq.(1) with $v = 1.86\times10^3$ m/s, $d = 1.85\times10^{-6}$ m yields $\tau$=3.2 ns and $\lambda_N = 2.8$ μm, respectively.

## IV. Discussion
### (a) Numerical calculation of spin accumulation voltage in nonmagnetic channels

To reveal the spin transport phenomena in the nondegenerate Si, a theoretical model of the spin accumulation voltages in a FE/NC/FE geometry, as schematically shown in Fig. 4(a), is constructed and discussed. Whereas the spin accumulation in nonmagnetic metal-based LSVs was established and discussed previously [36-42], the spin accumulation in nondegenerate semiconductors-based LSVs has not been discussed. To construct such a model, we start from the spin-drift-diffusion equation for nondegenerate semiconductors established by Yu and Flatte

[43, 44]. The physical parameters used in the numerical calculation are summarized in Table 1. An important component of their theory is that the spin carrier density, instead of the electrochemical potential, governs the equation (Eq. (2)) because the spin carriers (=charge carriers) are thermally activated in nondegenerate semiconductors. Hence,

$$\nabla^2 (n_\uparrow - n_\downarrow) - \frac{eE}{k_B T} \nabla (n_\uparrow - n_\downarrow) - \frac{(n_\uparrow - n_\downarrow)}{\lambda_N} = 0, \tag{2}$$

where $n_{\uparrow(\downarrow)}$ is carrier density of up (down) spin, $e(<0)$ is the electric charge, $E$ (>0) is the electric field in non-degenerate semiconductor, $k_B$ is the Boltzmann constant, $T$ is the temperature. The electrochemical potential of up (down) spins, $\mu_{\uparrow(\downarrow)}$, can be described as $\mu_{\uparrow(\downarrow)}(x_N) = -eEx_N + k_B T \ln\left(1 + \frac{\Delta n_{\uparrow(\downarrow)}}{n^0_{\uparrow(\downarrow)}}\right)$, where $n^0_{\uparrow(\downarrow)}$ is the electron density of the up (down) spins at equilibrium, $\Delta n_{\uparrow(\downarrow)}$ is the deviation of electron density of up (down) spin from the equilibrium and $x_N$ are the position in NC. Hence, the electrochemical potentials in FE1, FE2, NC3, NC4 and NC5 (see Fig. 4(a)), $\mu_{m\uparrow(\downarrow)}$ (m=1~5) are described as follows:

$$\mu_{1\uparrow(\downarrow)}(x_F) = -\frac{eJ}{\sigma_{F1}} x_F + (-)\mu_1(0) \frac{\sigma_{F1\downarrow(\uparrow)}}{\sigma_{F1}} e^{\frac{x_F}{\lambda_{F1}}} + R_{i1\uparrow(\downarrow)} e J^0_{1\uparrow(\downarrow)} + E_\mu \tag{3}$$

$$\mu_{2\uparrow(\downarrow)}(x_F) = -\frac{eJ}{\sigma_{F2}}(x_F - d) - \frac{eJ}{\sigma_N} d + (-)\mu_2(d) \frac{\sigma_{F2\downarrow(\uparrow)}}{\sigma_{F2}} e^{\frac{d-x_F}{\lambda_{F2}}} - R_{i2\uparrow(\downarrow)} e J^0_{2\uparrow(\downarrow)} - D_\mu \tag{4}$$

$$\mu_{3\uparrow(\downarrow)}(x_N) = k_B T \ln\left(1 + (-)\alpha_3(0) e^{-\frac{x_N}{\lambda_d}} + (-)\alpha_3(d) e^{\frac{x_N - d}{\lambda_u}}\right) - \frac{eJ}{\sigma_N} x_N \tag{5}$$

$$\mu_{4\uparrow(\downarrow)}(x_N) = k_B T \ln\left(1 + (-)\alpha_4(0) e^{-\frac{x_N}{\lambda_N}}\right) \tag{6}$$

$$\mu_{5\uparrow(\downarrow)}(x_N) = -\frac{eJ}{\sigma_N} d + (-)k_B T \ln\left(1 + \alpha_5(0) e^{\frac{d-x_N}{\lambda_N}}\right), \tag{7}$$

where $x_F$ is the position in FEs, $J$ is the charge current density, $\lambda_{F1(2)}$ is the spin diffusion length in FE1(2), $\lambda_{u(d)}$ are the upstream (downstream) spin transport length scale in NC, $R_{i1\uparrow(\downarrow)}$ and $R_{i2\uparrow(\downarrow)}$ are the spin-dependent interfacial resistance, $\mu_1(0)$ and $\mu_2(d)$ are the deviations in the electrochemical potential from the equilibrium in FE1 and FE2, respectively, and $\alpha_j(x_N)$ (j = 3, 4, 5) is $\frac{\Delta n_{\uparrow(\downarrow)}}{n^0_{\uparrow(\downarrow)}}$ at $x_N$. $\lambda_{u(d)}$ is expressed as $\left[-(+)\frac{|eE|}{2}\frac{\mu}{eD} + \sqrt{\left(\frac{|eE|}{2}\frac{\mu}{eD}\right)^2 + \frac{1}{\lambda_N^2}}\right]^{-1}$, where $\mu$ is the effective mobility. Assuming that the deviation of electrochemical potential in NC, $\mu_j(x_N)$, bears a linear relationship to $\frac{\Delta n_{\uparrow(\downarrow)}}{n^0_{\uparrow(\downarrow)}}$, $\mu_{j\uparrow(\downarrow)}(x_N)$ is expressed as follows:

$$\mu_{3\uparrow(\downarrow)}(x_N) = +(-)\frac{\mu_3^+}{2}e^{-\frac{x_N}{\lambda_d}} + (-)\frac{\mu_3^-}{2}e^{\frac{x_N-d}{\lambda_u}} - \frac{eJ}{\sigma_N}x_N \tag{8}$$

$$\mu_{4\uparrow(\downarrow)}(x_N) = +(-)\frac{\mu_4^-}{2}e^{\frac{x_N}{\lambda_N}} \tag{9}$$

$$\mu_{5\uparrow(\downarrow)}(x_N) = -\frac{eJ}{\sigma_N}d + (-)\frac{\mu_5^+}{2}e^{\frac{d-x_N}{\lambda_N}} . \tag{10}$$

Of note, Eqs. (8)-(10) have the same notation as those of nonmagnetic degenerate semiconductors and nonmagnetic metal channels. Finally, we obtained a spin-dependent voltage $V_{m\uparrow(\downarrow)} = \frac{\mu_{m\uparrow(\downarrow)}}{e}$ ($m$ =1~5) and current densities of up (down) spin $J_{m\uparrow(\downarrow)}$ as follows:

$$V_{1\uparrow(\downarrow)}(x_F) = -\frac{J}{\sigma_{F_1}}x + (-)V_1^- \frac{\sigma_{F_1\downarrow(\uparrow)}}{\sigma_{F_1}}e^{\frac{x_F}{\lambda_{F_1}}} + R_{i1\uparrow(\downarrow)}J_{1\uparrow(\downarrow)}^0 + E_V \tag{11}$$

$$J_{1\uparrow(\downarrow)}(x_F) = \frac{\sigma_{F_1\uparrow(\downarrow)}}{\sigma_{F_1}}J - (+)\frac{\sigma_{F_1\uparrow}\sigma_{F_1\downarrow}}{\sigma_{F_1}}\frac{1}{\lambda_{F_1}}e^{\frac{x_F}{\lambda_{F_1}}}V_1^- = +\frac{\sigma_{F_1\uparrow(\downarrow)}}{\sigma_{F_1}}J - (+)\frac{1}{R_{F_1}}e^{\frac{x_F}{\lambda_{F_1}}}V_1^- \tag{12}$$

$$V_{2\uparrow(\downarrow)}(x_F) = -\frac{J}{\sigma_{F_2}}(x_F - d) - \frac{J}{\sigma_N}d + (-)V_2^+ \frac{\sigma_{F_2\downarrow(\uparrow)}}{\sigma_{F_2}}e^{\frac{d-x_F}{\lambda_{F_2}}} - R_{i2\uparrow(\downarrow)}J_{2\uparrow(\downarrow)}^0 - D_V \tag{13}$$

$$J_{2\uparrow(\downarrow)}(x_F) = +\frac{\sigma_{F_2\uparrow(\downarrow)}}{\sigma_{F_2}}J + (-)\frac{1}{R_{F_2}}e^{\frac{d-x_F}{\lambda_{F_2}}}V_2^+ \tag{14}$$

$$V_{3\uparrow(\downarrow)}(x_N) = +(-)\frac{V_3^+}{2}e^{-\frac{x_N}{\lambda_d}} + (-)\frac{V_3^-}{2}e^{\frac{x_N-d}{\lambda_u}} - \frac{J}{\sigma_N}x_N \tag{15}$$

$$J_{3\uparrow(\downarrow)}(x_N) = +(-)\frac{1}{2R_d}V_3^+ e^{-\frac{x_N}{\lambda_d}} - (+)\frac{1}{2R_u}V_3^- e^{\frac{x_N-d}{\lambda_u}} + \frac{J}{2} \tag{16}$$

$$V_{4\uparrow(\downarrow)}(x_N) = +(-)\frac{V_4^-}{2}e^{\frac{x_N}{\lambda_N}} \tag{17}$$

$$J_{4\uparrow(\downarrow)}(x_N) = -(+)\frac{\frac{1}{2}\sigma_N}{\lambda_N}\frac{V_4^-}{2}e^{\frac{x_N}{\lambda_N}} = -(+)\frac{1}{2R_N}V_4^- e^{\frac{x_N}{\lambda_N}} \tag{18}$$

$$V_{5\uparrow(\downarrow)}(x_N) = -\frac{J}{\sigma_N}d + (-)\frac{V_5^+}{2}e^{\frac{d-x_N}{\lambda_N}} \tag{19}$$

$$J_{5\uparrow(\downarrow)}(x_N) = +(-)\frac{1}{2R_N}V_5^+ e^{\frac{d-x_N}{\lambda_N}} \tag{20}$$

where $R_{F1(2)} = \left(\frac{1}{\sigma_{F1(2)\uparrow}} + \frac{1}{\sigma_{F1(2)\downarrow}}\right)\lambda_{F1(2)}$, $2\left(\frac{1}{\sigma_N}\right)\lambda_N = R_N$ and $2\left(\frac{1}{\sigma_N}\right)\lambda_{u(d)} = R_{u(d)}$.

The boundary conditions of $V_{m\uparrow(\downarrow)}$ and $J_{m\uparrow(\downarrow)}$ at $x_N=x_F=0$ and $x_N=x_F=d$ are described as

follows:

At $x_N = x_F = 0$:

$$+(-)V_1^- \frac{\sigma_{F1\downarrow(\uparrow)}}{\sigma_{F1}} + R_{i1\uparrow(\downarrow)} J_{1\uparrow(\downarrow)}^0 + D_V = +(-)\frac{V_3^+}{2} + (-)\frac{V_3^-}{2}\eta_u = +(-)\frac{V_4^-}{2} \quad (21)$$

$$J_{1\uparrow(\downarrow)}^0 = +\frac{\sigma_{F1\uparrow(\downarrow)}}{\sigma_{F1}} J - (+)\frac{1}{R_F} V_1^- = +(-)\frac{1}{2R_d} V_3^+ - (+)\frac{1}{2R_u} V_3^- \eta_u + \frac{J}{2} + (-)\frac{1}{2R_N} V_4^- \quad (22)$$

At $x_N = x_F = d$:

$$+(-)V_2^- \frac{\sigma_{F2\downarrow(\uparrow)}}{\sigma_{F2}} - R_{i2\uparrow(\downarrow)} J_{2\uparrow(\downarrow)}^0 - E_V = +(-)\frac{V_3^+}{2}\eta_d + (-)\frac{V_3^-}{2} = +(-)\frac{V_5^+}{2} \quad (23)$$

$$J_{2\uparrow(\downarrow)}^0 = +\frac{\sigma_{F2\uparrow}}{\sigma_{F2}} J + (-)\frac{1}{R_{F2}} V_2^+ = +(-)\frac{1}{2R_d}\eta_d V_3^+ - (+)\frac{1}{2R_u} V_3^- + \frac{J}{2} - (+)\frac{1}{2R_N} V_5^+ \quad (24)$$

Finally, $E_V$, $D_V$, $V_3^+$, $V_3^-$ are obtained as follows:

$$E_V = + \left\{ \alpha_{F1} + (-\sigma_{F1\uparrow} R_{i1\uparrow} + \sigma_{F1\downarrow} R_{i1\downarrow}) \frac{q_d}{\sigma_{F1}} \right\} \frac{V_3^+}{2} + \left\{ \alpha_{F1} + (\sigma_{F1\uparrow} R_{i1\uparrow} - \sigma_{F1\downarrow} R_{i1\downarrow}) \frac{u_u}{\sigma_{F1}} \right\} \frac{V_3^-}{2} \eta_u - (\sigma_{F1\uparrow} R_{i1\uparrow} + \sigma_{F1\downarrow} R_{i1\downarrow}) \frac{J}{2\sigma_{F1}}$$

... (25)

$$D_V = + \left\{ -\alpha_{F2} - (\sigma_{F2\uparrow} R_{i2\uparrow} - \sigma_{F2\downarrow} R_{i2\downarrow}) \frac{u_d}{\sigma_{F2}} \right\} \frac{V_3^+}{2} \eta_d + \left\{ -\alpha_{F2} + (\sigma_{F2\uparrow} R_{i2\uparrow} - \sigma_{F2\downarrow} R_{i2\downarrow}) \frac{q_u}{\sigma_{F2}} \right\} \frac{V_3^-}{2} - (\sigma_{F2\uparrow} R_{i2\uparrow} + \sigma_{F2\downarrow} R_{i2\downarrow}) \frac{J}{2\sigma_{F2}}$$

... (26)

$$V_3^+ = \frac{\{-(Q_u+1)(R_{F1}\alpha_{F1} + R_{i1\uparrow} - R_{i1\downarrow}) + (U_u - 1)\eta_u (R_{F2}\alpha_{F2} + R_{i2\uparrow} - R_{i2\downarrow})\}\frac{J}{2}}{-(Q_d+1)(Q_u+1) + (U_u-1)(U_d-1)\eta_u\eta_d} \quad ... (27)$$

$$V_3^- = \frac{\{-(U_d-1)\eta_d(R_{F1}\alpha_{F1} + R_{i1\uparrow} - R_{i1\downarrow}) + (Q_d+1)(R_{F2}\alpha_{F2} + R_{i2\uparrow} - R_{i2\downarrow})\}\frac{J}{2}}{(U_u-1)(U_d-1)\eta_u\eta_d - (Q_d+1)(Q_u+1)} \quad ...(28)$$

, where $Q_d = \frac{1}{2}(R_F - R_{i1\uparrow} - R_{i1\downarrow})\left(\frac{R_N + R_d}{R_d R_N}\right)$, $Q_u = \frac{1}{2}(R_{F2} - R_{i2\uparrow} - R_{i2\downarrow})\left(\frac{R_N + R_u}{R_u R_N}\right)$, $U_d = \frac{1}{2}(R_{F2} - R_{i2\uparrow} - R_{i2\downarrow})\left(\frac{R_N - R_d}{R_d R_N}\right)$, $U_u = \frac{1}{2}(R_F - R_{i1\uparrow} - R_{i1\downarrow})\left(\frac{R_N - R_u}{R_u R_N}\right)$.

The spin accumulation voltage under the antiparallel configuration is also obtained by changing the polarity of $\alpha_{F2}$ and exchanging the subscripts of $\sigma_{F2\uparrow(\downarrow)}$ and $R_{i2\uparrow(\downarrow)}$. Here, we note that the above equations are applicable for LSVs with degenerate semiconductors and metal channels.

The spin-dependent voltage, $V_{3\uparrow(\downarrow)}(x_N)$ of up and down spins in the NC3 under parallel and antiparallel configurations calculated by Eqs. (15), (27) and (28) are shown in Figs. 4(b) and 4(e), where the voltage drop due to the electric field is subtracted. The solid and broken lines represent $V_{3\uparrow(\downarrow)}(x_N)$ under parallel and antiparallel configurations, respectively. Here, we assume $\sigma_N = 1 \times 10^4$ $\Omega^{-1} m^{-1}$, $d = 2$ μm, $\lambda_N = 1$ μm, $\lambda_F = 3$ nm, $R_{i1} = (R_{i1\uparrow}^{-1} + R_{i1\downarrow}^{-1})^{-1} = R_{i2} = 400$ Ω and $\alpha_{F1} = \alpha_{F2} = 0.1$. To simplify, we also assume that the spin polarization of the interface,

$\left(\frac{R_{i1\downarrow}-R_{i1\uparrow}}{R_{i1\downarrow}+R_{i1\uparrow}}\right) = \alpha_{i1}$ and $\left(\frac{R_{i2\downarrow}-R_{i2\uparrow}}{R_{i2\downarrow}+R_{i2\uparrow}}\right) = \alpha_{i2}$, is the same for $\alpha_{F1}$ and $\alpha_{F2}$, respectively [63-66]. First, we consider the spin accumulation without the spin drift effect (Fig. 4(b)). The direction of the accumulated spin generated from FE1 (spin extraction condition) and FE2 (spin injection condition) are opposite to each other under the parallel configuration, resulting in a small amount of spin accumulation. In contrast, under the antiparallel configuration, because the same spin are injected and extracted from FE1 and FE2, respectively, spin accumulation is enhanced. The features of the spin-dependent voltage are drastically changed when the spin drift effect is considered (Fig. 4(e)). Because the $\lambda_u$ of extracted spin from FE2 to FE1 is suppressed due to the upstream spin drift effect, spin accumulation in the Si channel adjacent to FE1 is almost insensitive to the magnetic configuration of FE2. The spin accumulation voltage obtained from L-3T MR was calculated from Eqs. (25)-(28). Two-dimensional color plots of $\Delta D_V$, i.e., the difference in $D_V$ between antiparallel and parallel configuration at the FE2 (under spin injection condition) and $\Delta E_V$ at the FE1 (under spin extraction condition), calculated without the spin drift effect, are shown in Figs. 4(c) and 4(d), respectively. The magnitude of $\Delta D_V$ and $\Delta E_V$ is increased with decreasing $\sigma_N$ and/or increasing $I$, which is a familiar characteristic in the conventional spin drift-diffusion model [36-42]. It is noted that $\Delta D_V$ and $\Delta E_V$ show the same value over the whole $\sigma_N$ and $I$ ranges, and this is inconsistent with the experimental results in Fig. 2(a). The calculated results that account for the spin drift effect are shown in Figs. 4(f) and 4(g). Compared to Fig. 4(d), the enhancement of $\Delta E_V$ is recognized in the high $I$ and low $\sigma_N$ region. At $I$ = 20 mA, $\sigma_N$ = 2000 $\Omega^{-1}$m$^{-1}$, $\Delta E_V$ in Fig. 4(g) is an order of magnitude higher than that in Fig. 4(d). In contrast, a significant reduction in $\Delta D_V$ occurs under the high $I$ and low $\sigma_N$ region in Fig. 4(f). The significant bias-current-polarity dependence is consistent with the experimental results in Fig. 2(a). Using, $\lambda_N$ = 2.8 μm, $\lambda_F$ = 3 nm, $d$ =1.85 μm, $R_{int.}$= $R_{int2}$= 750 Ω, and assuming $\alpha_{F1}$ =$\alpha_{F2}$=$\alpha_{i1}$=$\alpha_{i2}$, $\alpha_{F1}$ in Fig. 2(a) is estimated to be 0.085, a value consistent with that of the Fe/MgO/degenerate Si LSV [24, 26, 50]. In contrast, the conventional spin drift-diffusion model without spin drift effect yields $\alpha_{F1} \simeq$ 0.16, which is approximately twice as high as that estimated by the new model.

**(b) Spin accumulation voltage in metallic channel and degenerate Si channel**

The significance of Eqs. (25)-(28) is that the model allows a successful explanation of the large spin accumulation voltage in the nondegenerate Si-based LSV measured by the L-3T MR and its significant bias-current-polarity dependence. To corroborate the versatility of the model, L-3T MR on a degenerated Si-based and a Cu-based LSVs was also carried out. The device structure and results of the L-3T MR for a degenerated Si-based LSV are shown in Figs. 5(a), 5(b) and 5(c). Under the spin extraction condition (Fig. 5(b)), a clear rectangular signal with $\Delta E_V/I = 24$ mΩ ($\Delta E_V = 72$ μV) was detected. The L-3T MR under the spin injection condition after background subtraction is shown in Fig. 5(c) (the raw data are shown in Fig. 5(d)). Although a triangular hysteresis feature with resistance change of $\Delta D_V/I = 3$ mΩ ($\Delta D_V = 9$ μV) at ±300 Oe was obtained, it is mainly due to the spurious effect discussed in Fig. 2(a). However, in contrast to the nondegenerate Si-based LSV, a small resistance change of $\Delta D_V/I = 1$ mΩ ($\Delta D_V = 3$ μV) at ±100 Oe was also recognized, which is attributed to the spin accumulation, whereas $\Delta D_V$ is an order of magnitude smaller than $\Delta E_V$.

The device structure and results of the L-3T MR in the Cu-based LSV are shown in Figs. 6(a), 6(b) and 6(c). A nonlocal voltage signal at RT shown in Fig. 6(d) revealed that the magnetization reversal of the Py wires occurred at ±70 and ±240 Oe. The L-3T MR clearly showed a rectangular hysteresis feature both under spin injection and extraction conditions. It should be noted that $\Delta D_V$ (0.50 μV) is almost the same as $\Delta E_V$ (0.49 μV). Here, we focus on the ratio of $\Delta D_V$ to $\Delta E_V$. Theoretical and experimental values of $\Delta D_V/\Delta E_V$ for the nondegenerate Si-, the degenerate Si-, and the Cu-based LSVs are summarized in Table 2. The theoretical results of $\Delta D_V/\Delta E_V$ roughly correspond to the experimental ones, which demonstrates the usability of the new model in a wide range of $\sigma_N$.

$\Delta E_V$ as a function of $I$ in the nondegenerate Si-, the degenerate Si-, and the Cu-based LSV is displayed in Figs. 7(a)-7(c), respectively. $\Delta E_V$ of the nondegenerate-Si LSV (Fig. 7(a)) shows a superlinear and sublinear relationship below $4 \times 10^{-4}$ A and above $6 \times 10^{-4}$ A, respectively. A theoretical calculation is also plotted in the same figure. Although the theoretical calculation reproduces the superlinear feature, it is maintained even above $I = 6 \times 10^{-4}$ A. One possible origin of the sublinear feature is degradation of the spin transport properties e.g., spin polarization, spin lifetime, and so on, due to device heating [67, 68]. In contrast, in the degenerate Si- and Cu-based LSVs, a weak superlinear relationship and a linear relationship were obtained, respectively, which were well reproduced by the theoretical curves. Therefore, the nonlinear bias current dependence of $\Delta E_V$ provides additional evidence for the spin drift effect.

**(c) Features of the spin drift effect**

The spin drift effect modulates the spin transport length, resulting in a long-range spin transport of more than 21 μm at room temperature [34]. Due to a modulation of the spin transport length, the spin accumulation voltage under spin injection and extraction conditions is suppressed and enhanced, respectively, as discussed in sections III, IV(a), IV(b). Here, we discuss the advantages of such a redistribution of spin accumulation. One of the evident advantages is a reduction of the noise level. The standard deviations of the voltage measurement in the L-3T scheme in the nondegenerate-Si LSV were $4\times10^{-5}$ V and $2\times10^{-3}$ V under spin extraction and injection conditions, respectively. The noise level under the spin extraction condition was two orders of magnitude smaller than that under the spin injection condition. In the local two-terminal measurements, the standard deviation was $2\times10^{-3}$ V, which is comparable to that in the L-3T MR under the spin injection condition. This result indicates that whereas the spin accumulation voltage is mainly detected at FE under the spin extraction, the noise is mainly generated at FE under the spin injection. Spatial separation of the sources of noise and spin signal in the L-3T MR measurement is a great advantage for high-sensitivity measurements.

Hereafter, we focus on the spin drift effect on the two terminal magnetoresistance, which is a typical scheme for practical spin devices such as spin MOSFETs [16, 17]. Fig. 8 shows a comparison of the two-terminal local magnetoresistance, $\Delta R$, between the cases with and without the spin drift effect calculated with Eqs.(25)~(28). In the metallic spin channel, in which $\sigma_N=1\times10^7$ $\Omega^{-1}$m$^{-1}$ (Fig. 8(a)), the magnitude of $\Delta R$ is almost the same, indicating that spin drift effect is negligibly small. For $\sigma_N=1\times10^6$ $\Omega^{-1}$m$^{-1}$, the difference is recognized only at a higher $I$ region above 10 mA; however, the enhancement is only 6 % at 20 mA. The enhancement becomes more pronounced with decreasing channel conductivity and reaches 110 % at 1 mA and 580 % at 20 mA for $\sigma_N=1\times10^4$ $\Omega^{-1}$m$^{-1}$ (Fig. 8(d)). Therefore, the spin drift effect not only induces the redistribution of spin accumulation between spin source and drain in the spin MOSFET but also enhances the magnetoresistance. Taking advantage of the spin drift effect is the key for high-performance operation of the semiconductor-based spin devices.

**V Summary**

We investigated the spin-accumulation voltage measured by the local three-terminal magnetoresistance (L-3T MR) and compared the experimental and theoretical data. Although no clear spin-accumulation voltage was detected under the spin injection condition, a significantly large spin-accumulation voltage of more than 1.5 mV at 1 mA, i.e., a magnetoresistance of 1.5 $\Omega$, was obtained in the nondegenerate Si-based lateral spin valves under the spin extraction conditions. A spin drift-diffusion model that accounts for the spin drift effect successfully explained the large spin-accumulation voltage and significant bias-current-polarity dependence.

The model nicely reproduced output voltages from lateral spin valves with different conductivities in the range of more than three orders of magnitude, which corroborates the model constructed in this study. We also showed that the spin drift effect enhances magnetoresistance in the two terminal scheme, which is significant for the further development of high-performance spin MOSFET.

**Acknowledgements**

This research was supported in part by a Grant-in-Aid for Scientific Research from the Ministry of Education, Culture, Sports, Science and Technology (MEXT) of Japan (Innovative Area "Nano Spin Conversion Science" No. 26103003, "Grant-in-Aid for Scientific Research (A)" No. 25246019 and "Grant-in-Aid for Young Scientists (A)" No. 25709027).

**Additional information**

The authors declare no competing financial interests.

**Figure captions**

**Figure 1**

(a) Schematic of the silicon (Si)-based lateral spin valve (LSV). Two Fe ferromagnetic electrodes (FE1, FE2) are placed in contact with two Al nonmagnetic electrodes (NE1, NE2). A MgO barrier is used to realize a large spin accumulation in the Si channel. (b) A schematic of the electrochemical potential of up and down spins in a nonmagnetic channel (NC) and FEs. An electric field generates a charge current, $I$, from FE1 to FE2, resulting in spin injection and extraction at the NC/FE2 and FE1/NC interfaces, respectively. A spin accumulation in NC is generated due to the spin injection and extraction, resulting in a spin-dependent voltage drop, (spin injection: $\Delta D_V$ and spin extraction: $\Delta E_V$) at the interfaces, which induce two-terminal magnetoresistance effect.

**Figure 2**

(a) Local three-terminal magnetoresistances (L-3T MR) in the nondegenerate Si-based LSV measured at room temperature. The currents of the upper and lower panels were +1 mA (spin extract condition) and −1 mA (spin injection condition), respectively. The distance, $d$, between FE1 and FE2 in Fig. 1(a) was 1.85 μm. The blue and red circles show the results obtained under different magnetic field sweep directions. (b) Hanle effect signals measured in the L-3T MR scheme at RT, where the magnetic field direction was perpendicular to the plane. The current was +1 mA. The raw data under parallel (red) and antiparallel (blue) configurations are shown in the inset and the difference in the Hanle signal between antiparallel and parallel configurations is shown in the main panel. The fit with Eq. (1) is shown by the black solid line.

**Figure 3**

L-3T MRs of two different devices B and C measured at +1 mA. The distance, $d$, between FE1 and FE2 of the devices B and C were 2.00 and 2.75 μm, respectively.

**Table 1**

Summary of the physical parameters used in the numerical calculation.

**Figure 4**

(a) A schematic of the Si-based LSV for calculation of spin accumulation voltage. The spin

dependent voltage $V_{3\uparrow(\downarrow)}(x_F)$ of up (red) and down (blue) spins under parallel (solid line) and anti-parallel (broken line) configurations calculated using Eqs. (15), (27) and (28), (b) without and (e) with the spin drift effect. Spin accumulation voltage detected at the FE1 (spin injection condition), $\text{Log}(\Delta D_V)$, and FE2 (spin extraction condition), $\text{Log}(\Delta E_V)$, calculated with Eqs. (25)-(28) with (c), (d) and without (f), (g) the spin drift effect. $\text{Log}(\Delta D_V)$ calculated to be less than −12 is plotted in the same color (purple).

**Figure 5**

(a) Schematic of degenerate Si-based LSV. L-3T MR measured at (b) +3 mA (spin extraction condition, $E_v/I$) and (c) −3 mA (spin injection condition, $D_v/I$). (b) The raw data of L-3T MR at −3 mA. All measurements were conducted at RT.

**Figure 6**

(a) Schematic of Cu-based LSV. L-3T MR signals in Cu-based LSV measured at (b) +2 mA ($E_v/I$) and (c) −2 mA ($D_v/I$). (c) Nonlocal voltage signal measured at −3 mA. All measurements were conducted at RT.

**Table 2**

Summary of experimentally obtained $\Delta D_V$, $\Delta E_V$ and $\Delta D_V/\Delta E_V$ and theoretically obtained $\Delta D_V/\Delta E_V$ for nondegenerate Si-, degenerate Si-, Cu-based LSVs.

**Figure 7**

Bias current dependence of $\Delta E_V$ of (a) nondegenerate Si-, (b) degenerate Si- and (b) Cu-based LSV measured at RT. Solid lines are the theoretical calculations with Eqs. (25)-(28).

**Figure 8**

Bias current dependence of two-terminal magnetoresistance calculated with Eqs. (25)-(28) both with (red) and without (blue) the spin drift effect. The conductivities of the channel are (a) $1\times10^7$, (b) $1\times10^6$ (c) $1\times10^5$, and (d) $1\times10^4$ $\Omega^{-1}\text{m}^{-1}$, respectively.

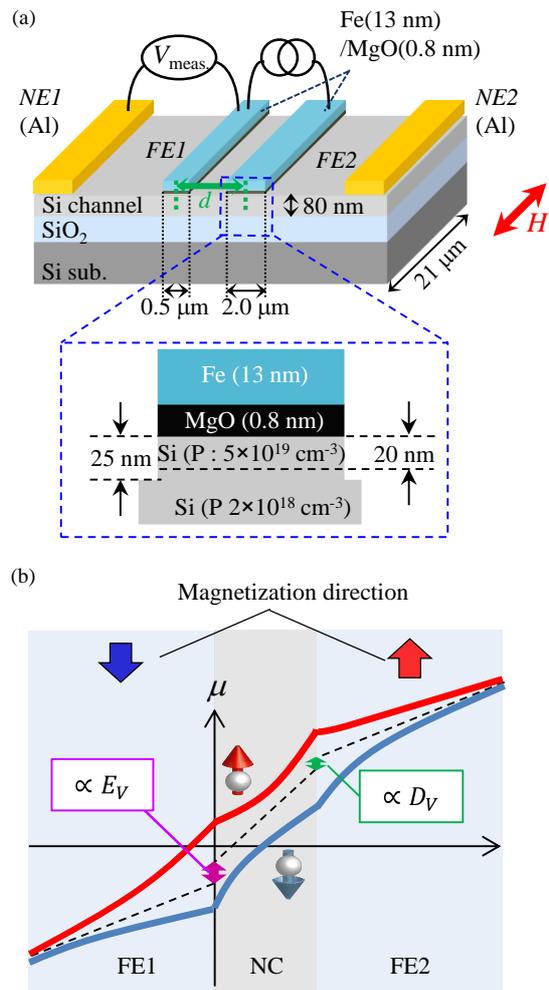

Fig 1 Tahara et al.

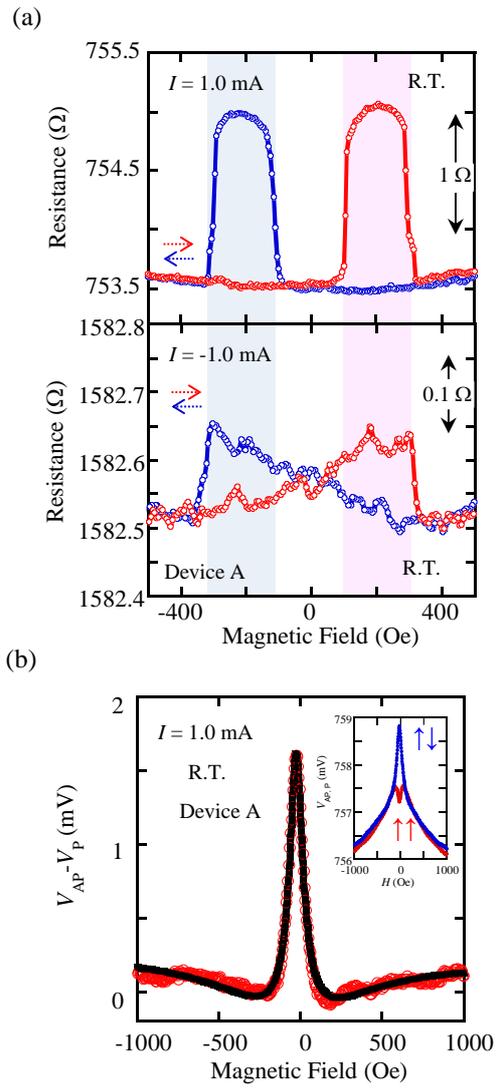

Fig 2 Tahara et al.

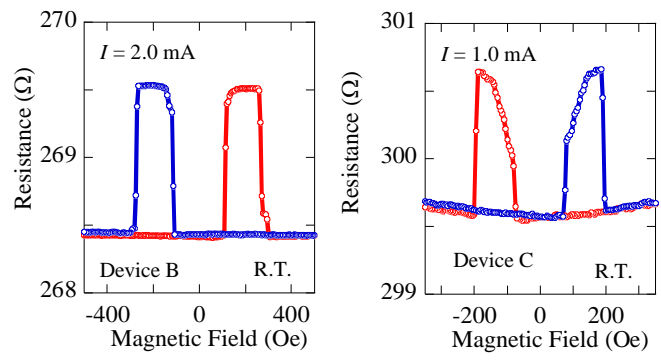

Fig. 3 Tahara et al.

# Table 1 Tahara et al

| Spin dependent voltage of up (down) spin | Charge current density | Charge current density of up (down) spin | Charge current density of up (down) spin at $x_F=0$, $d$ |
|---|---|---|---|
| $V_{m\uparrow(\downarrow)}$ (m=1~5) | $J$ | $J_{m\uparrow(\downarrow)}$ (m=1~5) | $J^0_{1\uparrow(\downarrow)}, J^0_{2\uparrow(\downarrow)}$ |

| Conductivity of FE1, FE2 | Conductivity of up (down) spin in FE1, FE2 | Conductivity of NC | Spin dependent resistance of up (down) spin at FE1/NC, FE2/NC interfaces |
|---|---|---|---|
| $\sigma_{F1}, \sigma_{F2}$ | $\sigma_{F1\uparrow(\downarrow)}, \sigma_{F2\uparrow(\downarrow)}$ | $\sigma_N$ | $R_{i1\uparrow(\downarrow)}, R_{i2\uparrow(\downarrow)}$ |

| Spin polarization of FE1, FE2 | Spin diffusion length in FE1, FE2 | Spin diffusion length in NC | Upstream, downstream spin transport length scale in NC |
|---|---|---|---|
| $\alpha_{F1}, \alpha_{F2}$ | $\lambda_{F1}, \lambda_{F2}$ | $\lambda_N$ | $\lambda_u, \lambda_d$ |

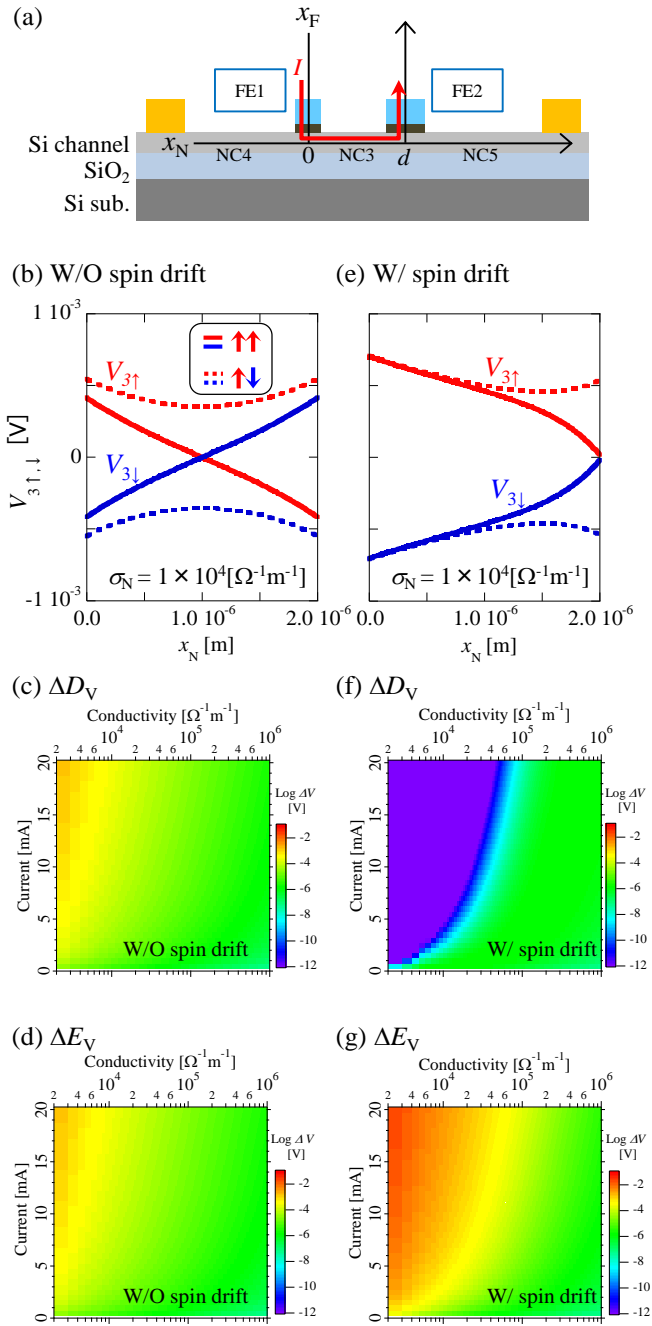

Fig. 4 Tahara et al.

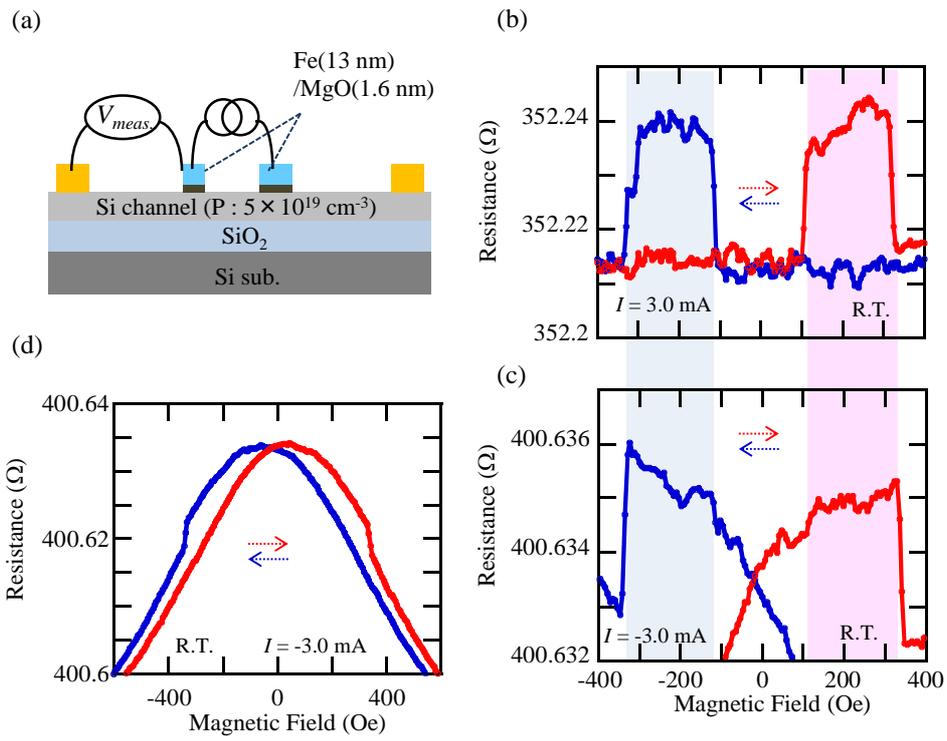

Fig 5 Tahara et al.

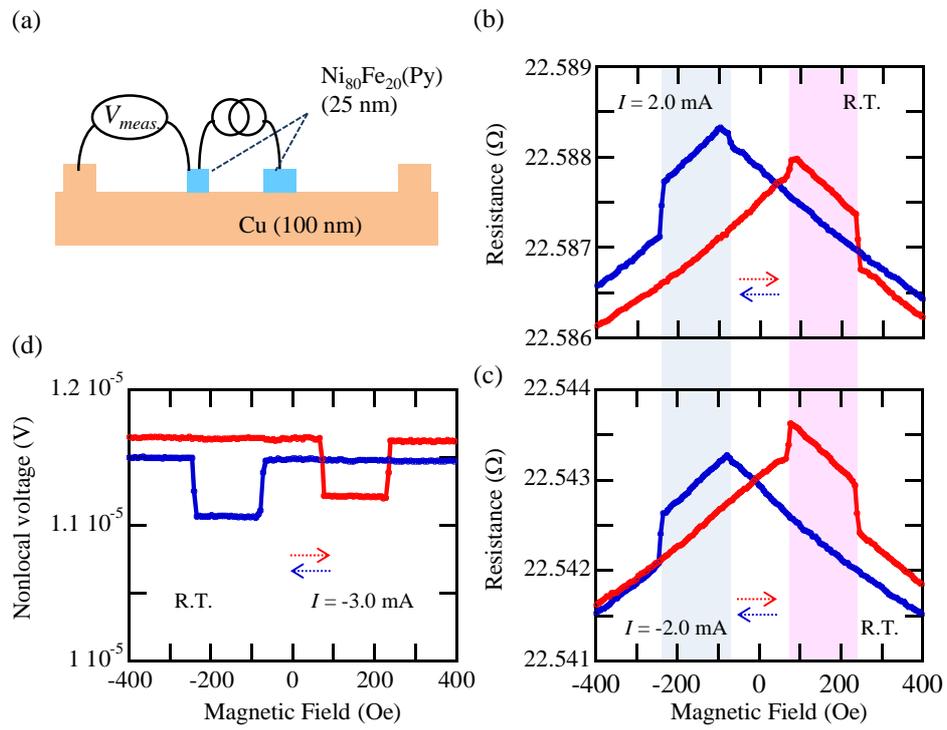

Fig 6 Tahara et al.

Table 2 Tahara et al

|  | Conductivity of the channel [$\Omega^{-1}$m$^{-1}$] | $\Delta D_V$ (Experiment) [V] | $\Delta E_V$ (Experiment) [V] | $\Delta D_V/\Delta E_V$ (Experiment) | $\Delta D_V/\Delta E_V$ (Calculation) |
|---|---|---|---|---|---|
| Nondegenerate Si ($I = \pm 1.0$ mA) | $2.3 \times 10^3$ | $< 10^{-5}$ | $1.5 \times 10^{-3}$ | $< 6.7 \times 10^{-3}$ | $7.0 \times 10^{-8}$ |
| Degenerate Si ($I = \pm 3.0$ mA) | $6.3 \times 10^4$ | $3.0 \times 10^{-6}$ | $7.2 \times 10^{-5}$ | $4.2 \times 10^{-2}$ | $5.0 \times 10^{-1}$ |
| Metal ($I = \pm 2.0$ mA) | $4.8 \times 10^7$ | $5.0 \times 10^{-7}$ | $4.9 \times 10^{-7}$ | $1.0 \times 10^0$ | $1.0 \times 10^0$ |

(a) Nondegenerate Si  (b) Degenerate Si  (c) Cu

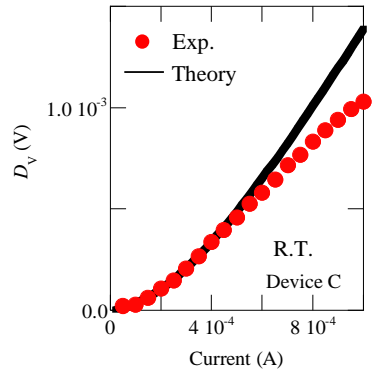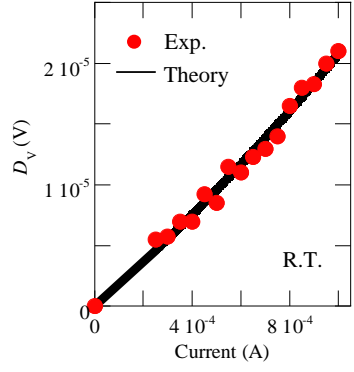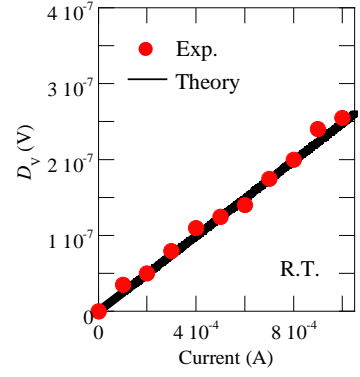

Fig. 7 Tahara et al.

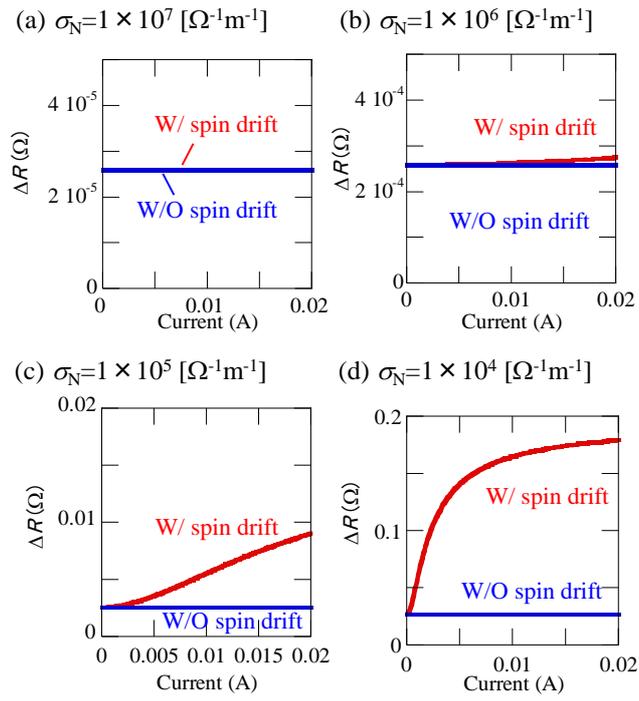

Fig. 8 Tahara et al.